\documentclass{article}

\usepackage[preprint]{Styles/neurips_2025}

\usepackage[utf8]{inputenc}
\usepackage[T1]{fontenc}
\usepackage{hyperref}
\usepackage{url}
\usepackage{booktabs}
\usepackage{amsfonts}
\usepackage{amsmath}
\usepackage{nicefrac}
\usepackage{microtype}
\usepackage{graphicx}
\usepackage{xcolor}
\graphicspath{{Images/}{./}}
\usepackage{ulem}

\makeatletter
\renewcommand{\@noticestring}{Conference on Physics and AI at Stanford University (PAI 2026).}
\makeatother
\title{Learning Transverse Momentum Distributions from Raw Scattering Events via Conditional Diffusion}

\author{%
  \textbf{Jitao Xu$^{1}$ \quad Christina Cocuzza$^{2}$ \quad Kevin Braga$^{2}$} \\[2pt]
  \textbf{Daniel Lersch$^{3}$ \quad Nobuo Sato$^{3}$ \quad Yaohang Li$^{1}$} \\[4pt]
  $^{1}$Department of Computer Science, Old Dominion University \\
  $^{2}$Department of Physics, William \& Mary \\
  $^{3}$Thomas Jefferson National Accelerator Facility%
}

\begin{document}

\maketitle

\begin{abstract}
Extracting transverse momentum dependent parton distribution functions (TMD PDFs) from semi-inclusive deep inelastic scattering (SIDIS) data is a central goal of the nucleon structure program at Jefferson Lab and the future Electron-Ion Collider. Traditional extraction methods rely on parameterized functional forms and iterative fitting, which can limit the flexibility of the resulting distributions and make uncertainty quantification cumbersome. We present a conditional diffusion model that learns to map raw SIDIS event kinematics directly to TMD PDFs, bypassing explicit functional assumptions. Evaluated on simulated SIDIS data at CLAS12 kinematics, the model recovers the underlying TMD with informative uncertainties that narrow steadily with increasing event statistics, and produces reliable estimates even with as few as 1{,}000 conditioning events, a statistics-limited regime directly relevant to ongoing and planned experiments.
\end{abstract}

\section{Introduction}

Transverse momentum dependent parton distribution functions (TMD PDFs) describe the joint distribution of quarks and gluons in longitudinal momentum fraction $x$ and transverse momentum $\boldsymbol{k}_T$ inside the proton, providing a three-dimensional picture of nucleon structure in momentum space~\citep{collins2011foundations}. Semi-inclusive deep inelastic scattering (SIDIS) is one of the primary channels for accessing TMDs experimentally. With the CLAS12 detector at Jefferson Lab already collecting high-statistics SIDIS data~\citep{burkert2020clas12} and the Electron-Ion Collider on the horizon~\citep{abdul2022eic}, the volume and kinematic reach of available measurements will grow substantially over the coming decade.
 
Standard TMD extractions parameterize the non-perturbative component of the distribution, then fit the parameters to binned cross-section data by $\chi^2$ minimization~\citep{collins2011foundations}. This strategy has produced reliable global fits, but the results depend on the chosen functional form, and propagating uncertainties through the full theory chain, where Sudakov evolution, matching corrections, and Fourier--Bessel transforms, remains expensive.
 
Motivated by these limitations, we explore an alternative extraction strategy based on conditional diffusion models~\citep{ho2020denoising,alghamdi2025eventlevel}. The idea is to learn the map from raw SIDIS event kinematics directly to the TMD, bypassing explicit parameterization. Specifically, the model represents the TMD on a discretized $(x, b_T)$ grid (here $b_T$ is the Fourier conjugate of $\boldsymbol{k}_T$; the two carry the same information) and learns to denoise it conditioned on a variable-size set of observed events, which are encoded using a permutation-invariant PointNet-Pool architecture~\citep{qi2017pointnet} followed by a transformer encoder. Because the diffusion process is inherently stochastic, posterior uncertainty is obtained simply by drawing multiple samples, with no additional inference step required.
 
We test the framework on simulated SIDIS data at CLAS12 kinematics. The extracted TMDs agree well with the ground truth, and the uncertainty bands narrow as the number of conditioning events increases, consistent with the expected statistical scaling.

\section{Method}
\begin{figure}[t]
  \centering
  \includegraphics[width=0.80\textwidth]{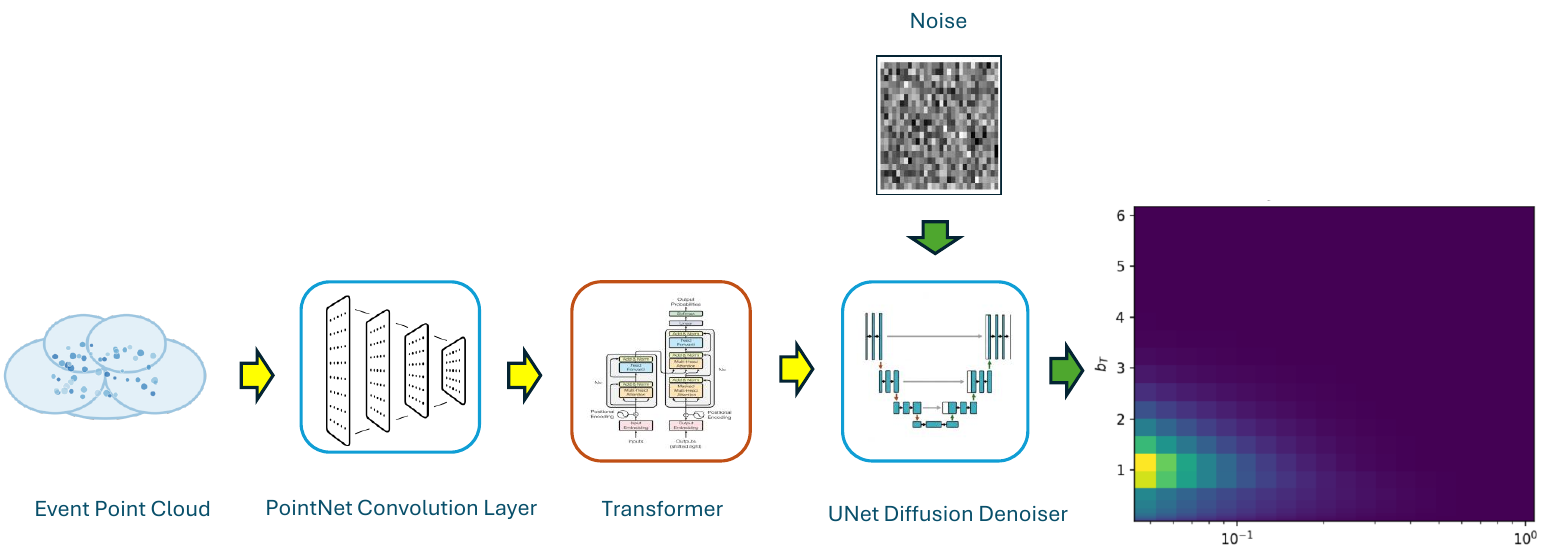}
  \caption{Overview of the proposed framework. A variable-size set of raw SIDIS events is first mapped to per-point features by a shared PointNet convolution layer, then aggregated through six-way global pooling and refined by a lightweight Transformer encoder to produce a fixed-dimensional conditioning vector. This vector, together with a sinusoidal timestep embedding, guides a 2D UNet that iteratively denoises Gaussian noise into a predicted TMD image $f(x, b_T)$.}
  \label{fig:architecture}
\end{figure}
\paragraph{Physics pipeline.}
The unpolarized SIDIS differential cross section at leading order is computed from the input TMD PDF $f(x, b_T)$ through a chain of physical operations~\citep{collins2011foundations,collins1985tmd}: (i) perturbative Sudakov evolution from the initial scale to $f(x, Q^2, b_T)$, (ii) convolution with the TMD fragmentation function $D(z,Q^2,b_T)$, taken from a prior extraction and held fixed throughout, and (iii) Fourier--Bessel transform from $b_T$-space to transverse momentum $q_T$-space, yielding the leading-order structure function.
Events $(x, Q^2, z, q_T, \phi)$ are then sampled from the resulting cross section via an inverse transform sampler.

To enable efficient evaluation and learning of the SIDIS cross section, we adopt a discretized representation of the underlying TMDs using a localized basis expansion. Specifically, functions such as $\tilde{f}(x,b_T,Q^2)$ and $\tilde{D}(z,b_T,Q^2)$ are expressed as linear combinations of basis functions defined on a grid in $(x, Q^2, b_T)$ space. This representation allows all subsequent operations in the SIDIS pipeline—including evolution, convolution, and Fourier–Bessel transforms—to act linearly on the basis functions rather than directly on the unknown coefficients. As a result, computationally expensive operators can be precomputed and reused, reducing the full pipeline to a sequence of tensor contractions in coefficient space. This structure is particularly well-suited for generative modeling, where the diffusion model learns to produce physically consistent coefficient configurations that are subsequently mapped to observable distributions through the fixed physics pipeline. Importantly, because the diffusion model interfaces with the theory solely through simulated events, the pipeline is not required to be differentiable or written in the same programming language \citep{braga2025toward}, which makes the framework readily applicable to other physics pipelines.

\paragraph{Conditional diffusion model.}
The overall model architecture is illustrated in Figure~\ref{fig:architecture}.
We employ a DDPM~\citep{ho2020denoising} with a 2D UNet backbone~\citep{ronneberger2015unet} to model the distribution of TMD images.
The UNet operates on $(20 \times 30)$ images with base channel width 96 and multipliers $(1, 2, 4)$, using a cosine noise schedule~\citep{nichol2021improved} with $T = 100$ diffusion steps.
The input TMD is transformed via $\log(1 + f)$ to compress the dynamic range before diffusion.
During training, a clean TMD image $x_0$ is corrupted by Gaussian noise according to
\begin{equation}
  x_t = \sqrt{\bar\alpha_t}\, x_0 + \sqrt{1 - \bar\alpha_t}\,\boldsymbol{\epsilon}, \quad \boldsymbol{\epsilon} \sim \mathcal{N}(\mathbf{0}, \mathbf{I}),
\end{equation}
where $\bar\alpha_t = \prod_{s=1}^{t} (1 - \beta_s)$ is the cumulative noise schedule.
The model $\boldsymbol{\epsilon}_\theta$ is trained to predict the added noise by minimizing
\begin{equation}
  \mathcal{L} = \mathbb{E}_{x_0,\, t,\, \boldsymbol{\epsilon}} \bigl[\|\boldsymbol{\epsilon}_\theta(x_t, t, c) - \boldsymbol{\epsilon}\|^2\bigr],
\end{equation}
where $c$ denotes the conditioning embedding from the event encoder.
At inference, new TMD samples are generated by iteratively denoising from $x_T \sim \mathcal{N}(\mathbf{0}, \mathbf{I})$.

\paragraph{Event conditioning.}
To condition on a variable-size set of $N_\text{ev}$ observed events, we use a PointNet-Pool encoder~\citep{qi2017pointnet}.
Each 5D event is processed by shared 1D convolutions ($5 \to 64 \to 128 \to 128$), followed by six-way global pooling that produces a fixed 641-dimensional summary vector:
\begin{equation}
  h_{\mathrm{pool}} = \bigl[\max_i h_i,\; \mathrm{mean}_i h_i,\; \min_i h_i,\; \mathrm{std}_i h_i,\; \mathrm{std}_i h_i / \sqrt{N},\; \log N \bigr],
\end{equation}
where $h_i$ are the per-point features and $N$ is the number of events. This vector is projected to 128 dimensions via a residual MLP.
The event embedding is added to the sinusoidal timestep embedding and injected into each UNet residual block.

\section{Experiments}

\subsection{Setup}
\begin{figure}[htbp]
  \centering
  \includegraphics[width=0.60\textwidth]{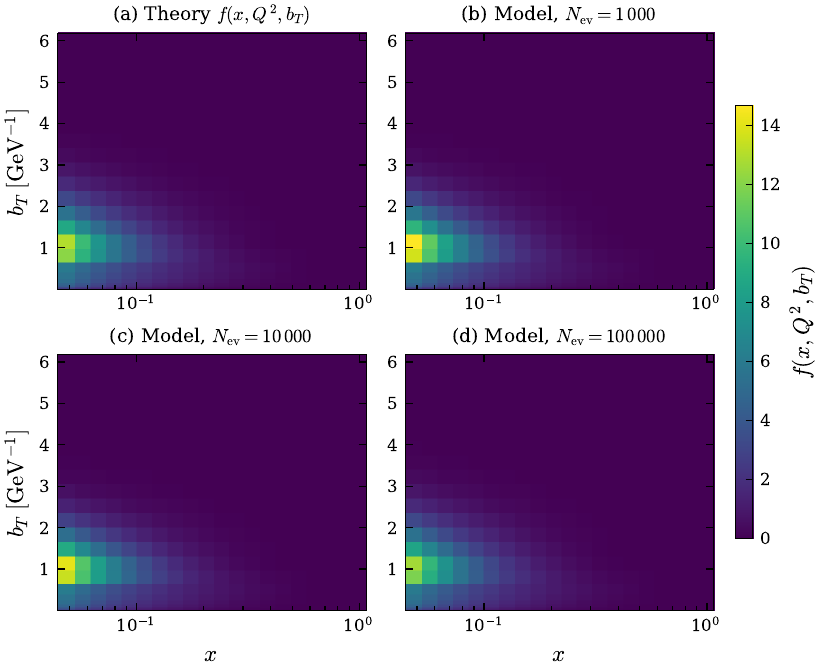}
  \caption{Evolved TMD $f(x, Q^2, b_T)$ on the full $(x, b_T)$ grid at a representative $Q^2$ slice. (a)~Theory ground truth. (b--d)~Diffusion model predictions (ensemble mean over 128 noise replicas) conditioned on $N_\text{ev} = 1{,}000$, $10{,}000$, and $100{,}000$ events, respectively.}
  \label{fig:tmd_heatmap}
\end{figure}
We evaluate the conditional diffusion model on SIDIS at $E_\text{beam} = 11$~GeV, matching CLAS12 kinematics at Jefferson Lab~\citep{burkert2020clas12}.
The TMD PDF is parameterized on a $20 \times 30$ grid in $(x, b_T)$ with $x \in [0.048, 1.0]$ (log-spaced) and $b_T \in [0.001, 6.0]$~GeV$^{-1}$.
The SIDIS theory module evolves the input TMD through perturbative Sudakov resummation, operator product expansion (OPE) corrections, and non-perturbative modeling, then computes differential cross section via Fourier transform from $b_T$-space to $q_T$-space.
Events are sampled from the resulting cross section using an inverse transform sampler.

Training uses a pre-generated parametric dataset of 50{,}000 TMD images with randomized non-perturbative parameters, optimized with Adam~\citep{kingma2015adam} (learning rate $5 \times 10^{-5}$, batch size 128) for 1{,}000 epochs.
To study the dependence on the number of conditioning events, we train three models with $N_\text{ev} \in \{1{,}000,\ 10{,}000,\ 100{,}000\}$, all other hyperparameters held fixed.

\subsection{Results}

Figure~\ref{fig:tmd_heatmap} shows the evolved TMD $f(x, Q^2, b_T)$ on the full $(x, b_T)$ grid, comparing the theory ground truth to the model predictions at each conditioning event count. The model progressively recovers the two-dimensional structure of the distribution as $N_\text{ev}$ increases.
\begin{figure}[htbp]
  \centering
  \includegraphics[width=0.65\textwidth]{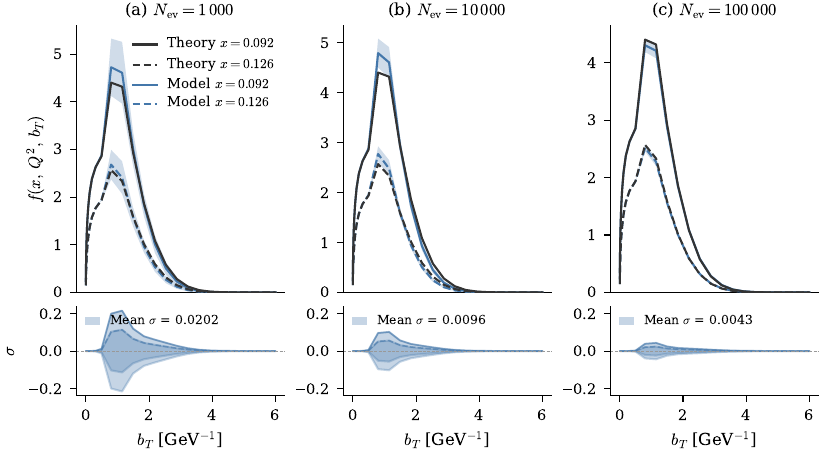}
  \caption{This figure shows one-dimensional slices of the two-dimensional TMD in Figure~\ref{fig:tmd_heatmap}, at fixed $x$ as a function of $b_T$. Evolved TMD slices at epoch 1{,}000 for three conditioning event counts. Dark gray: theory ground truth. Blue line and band: diffusion model mean $\pm\, 3\sigma$ over 128 noise replicas. Bottom panels: absolute model standard deviation $\sigma$. As $N_\text{ev}$ increases, the prediction converges to the true TMD with narrowing uncertainty.}
  \label{fig:tmd_slices}
\end{figure}

Figure~\ref{fig:tmd_slices} shows evolved TMD slices $f(x, Q^2, b_T)$ vs.\ $b_T$ at two representative $x$ values, comparing the theory ground truth (dark gray) to the diffusion model prediction (blue) with $\pm 3\sigma$ model uncertainty bands from 128 independent diffusion samples.

We summarize the spread of the diffusion posterior by $\langle\sigma\rangle$, the standard deviation across 128 independent samples averaged over all $(x, b_T)$ grid points.
With $N_\text{ev} = 1{,}000$ events, the model captures the overall shape of the TMD but exhibits a broad uncertainty band ($\langle\sigma\rangle \approx 0.020$), reflecting the limited statistical information in a small event sample.
At $N_\text{ev} = 10{,}000$, the mean closely tracks the theory curve across the full $b_T$ range with uncertainty narrowing to $\langle\sigma\rangle \approx 0.010$.
At $N_\text{ev} = 100{,}000$, model and theory are nearly indistinguishable, with $\langle\sigma\rangle \approx 0.004$.

This scaling behavior is physically expected: the statistical precision of the conditioning events improves as ${\sim}1/\sqrt{N_\text{ev}}$, and the diffusion model faithfully translates this into correspondingly tighter posterior predictions.
Crucially, even at $N_\text{ev} = 1{,}000$---an event count accessible in sparse kinematic bins or early-phase experiments---the model produces a reasonable TMD estimate with informative uncertainty.
This demonstrates that the conditional diffusion framework can operate in data-scarce regimes relevant to upcoming EIC measurements~\citep{abdul2022eic}, where many $(x, Q^2)$ bins will have limited statistics.

\section{Discussion and Conclusion}

We have demonstrated that a conditional DDPM can extract TMD PDFs from raw SIDIS event kinematics with informative uncertainties.
The PointNet-Pool encoder provides an effective, permutation-invariant interface between variable-size event sets and the diffusion model, enabling natural uncertainty quantification through repeated stochastic sampling.
The model's performance scales gracefully with the number of conditioning events, producing informative posteriors even in low-statistics regimes. 

Several directions remain for future work: extending to next-to-leading order (NLO) theory, incorporating polarized observables, training on real experimental data from CLAS12, scaling to large-scale multi-GPU training, and applying the framework to projected Electron-Ion Collider pseudo-data with realistic detector effects.
\begin{ack}
This work is partially supported by the U.S.\ Department of Energy, Office of Science, Office of Nuclear Physics, Office of Advanced Scientific Computing Research through the Scientific Discovery through Advanced Computing (SciDAC) program, under contracts DE-AC02-06CH11357, DE-AC05-06OR23177, and DE-SC0023472.
\end{ack}

\bibliographystyle{unsrtnat}
\bibliography{references}

\end{document}